\documentclass[12pt]{iopart}
\usepackage{graphicx,amstext,amssymb}
\usepackage{hyperref}
\usepackage{color}
\usepackage{multirow}

\begin{document}

\title{Symmetry restoration by pricing in a duopoly of perishable goods}
\author{Su Do Yi$^1$ Seung Ki Baek$^1$$^\ast$, Guillaume Chevereau$^2$, and Eric Bertin$^{3,4}$$^\dagger$}
\address{$^1$Department of Physics, Pukyong National University, Busan 48513, Korea}
\address{$^2$Laboratoire de Physique, ENS Lyon and CNRS, F-69007 Lyon, France}
\address{$^3$Universit\'e Grenoble Alpes, LIPHY, F-38000 Grenoble, France}
\address{$^4$CNRS, LIPHY, F-38000 Grenoble, France}
\ead{$^\ast$seungki@pknu.ac.kr}
\ead{$^\dagger$eric.bertin@ujf-grenoble.fr}

\begin{abstract}
Competition is a main tenet of economics, and the reason is that a perfectly
competitive equilibrium is Pareto-efficient in the absence of externalities
and public goods. Whether a product is selected in a market crucially relates
to its competitiveness, but the selection in turn affects the landscape of
competition. Such a feedback mechanism has been illustrated in a duopoly
model by Lambert et al., in which a buyer's satisfaction is updated depending on
the {\em freshness} of a purchased product. The probability for buyer $n$ to
select seller $i$ is assumed to be $p_{n,i} \propto e^{ S_{n,i}/T}$, where
$S_{n,i}$ is the buyer's satisfaction and $T$ is an effective temperature to
introduce stochasticity. If $T$ decreases below a critical point $T_c$, the
system undergoes a transition from a symmetric phase to an asymmetric one, in
which only one of the two sellers is selected.  In this work, we extend the
model by incorporating a simple price system.  By considering a {\em greed
factor} $g$ to control how the satisfaction depends on the price, we argue the
existence of an oscillatory phase in addition to the symmetric and asymmetric
ones in the $(T,g)$ plane, and estimate the phase boundaries through mean-field
approximations. The analytic results show that the market preserves the
inherent symmetry between the sellers for lower $T$ in the presence of the
price system, which is confirmed by our numerical simulations.
\end{abstract}


\maketitle
\section{Introduction}

Socio-economic systems have attracted attention of physicists due to their
inherent dynamic complexity~\cite{econophysics1,econophysics2,Bouchaud}.
Simple interactions between social agents are known to give rise to non-trivial
collective patterns like, e.g., segregation \cite{Schelling,Grauwin,Nadal},
opinion \cite{Barrat,Jensen1,Jensen2} or language dynamics \cite{Castellano},
and crowd behavior \cite{Theraulaz,Appert}. Focusing more specifically on
economic aspects, the interaction of sellers and buyers in a market can be
viewed as a dynamical process. For instance, taking space into account and
including transportation costs in the agents' utility leads to the well-known
Hotelling model \cite{Hotelling,Larralde}, in which stores try to find the
optimal location to maximize their profit.

Here we consider a different problem, neglecting spatial aspects but taking into
account a finite lifetime of products, leading to a potentially rich dynamics.
Buyers and sellers change their respective states upon every purchase:
Namely, each buyer evaluates the sellers based on the
purchased products, and each buyer also updates the list of products in
stock. Under certain condition, their interaction may form a positive
feedback loop in such a way
that a seller, if selected by a buyer, becomes more likely to be selected
in the future. Such a mechanism has been proposed and analyzed in detail
by Lambert et al.~\cite{Lambert2011}: They have considered two sellers dealing
with perishable goods, so that a seller can replace existing products
with fresh ones when buyers continuously make purchases from the seller.
As a result of this positive feedback, the symmetry among the sellers gets
broken spontaneously, leading to a virtual monopoly. Lambert et al. have
analytically identified a critical threshold for this phenomenon, when the
control parameter is the degree of randomness in buyers' choices.

However, their duopoly model in Ref.~\cite{Lambert2011} does not contain
any price system, and it is an interesting question whether the market can
restore the symmetry when the sellers are allowed to charge different prices. If
the price declines as time goes by, and if buyers are sensitive to the price as
well as freshness, it is indeed plausible that the market can better resist
the tendency toward a monopoly. We thus extend the model of
Ref.~\cite{Lambert2011} by including a price mechanism, which defines the
price of a product as a function of its freshness, and check to which extent
this prescription stabilizes the market.

In this paper, we show that the symmetry is recovered at a lower degree of
randomness in buyers by including a price system coupled to the freshness of
perishable goods. The threshold is estimated on a mean-field level in the
sense that we neglect fluctuations among buyers. We also find the possibility of
a third phase, in which buyers seesaw between the two sellers. The period of
this seesaw motion is estimated approximately by assuming homogeneity of the
products in stock. These analytic results are consistent with a phase diagram
obtained numerically.

This work is organized as follow: We explain our model in Sec.~\ref{model}.
In Sec.~\ref{subevol}, we analyze its time evolution under the assumption that
buyers remember the past for a long time. The threshold for symmetry breaking is
given by examining stationary states in Sec.~\ref{substat}. Section~\ref{subosc}
explains the origin of the oscillatory phase and gives an approximate estimate
of the period. We briefly discuss implications of our findings in
Sec.~\ref{discuss} and then conclude this work.

\section{Model}
\label{model}
Suppose that two sellers are competing to sell products of the same kind.
Each seller has $N_p$ products in stock, and each of them has its own age
$\tau$.
As soon as a product is sold, it is replaced by a new product with $\tau=0$ to
conserve the number of products on the market all the time.
As in Ref.~\cite{Lambert2011}, our assumption is that the products are
perishable, so that the freshness changes with $\tau$ in the following
functional way:
\begin{equation}
h(\tau)=e^{-\tau/\tau_1},
\label{eq.fresh}
\end{equation}
where $\tau_1$ is the characteristic time scale for aging.
It is a common marketing strategy to lower the prices of shopsoiled products.
We may therefore assume that each seller can choose a price policy, parametrized
by $h_c$: This parameter means the characteristic freshness for a markdown. If
$h_c$ is low, the price will not drop down until the product becomes very old.
Specifically, we set
\begin{equation}
x(h(\tau)) = x(\tau)=1-\exp \left[-\frac{h(\tau)}{h_c} \right],
\label{eq.price}
\end{equation}
which is a decreasing function of $\tau$, bounded between zero and
$1-e^{-h_c^{-1}}$.
On the other hand, we have $N_a$ buyers.
Let $n \in \{1,...,N_a\}$ be an index to denote each buyer.
If buyer $n$'s satisfaction from the $i$th seller is denoted by
$S_{n,i}$, the probability to choose this seller is given as
\begin{equation}
p_{n,i} = Z_n^{-1} e^{ S_{n,i}/T},
\label{eq.log}
\end{equation}
where $T$ is an effective temperature to control the degree of stochasticity
in making the decision, and $Z_n \equiv e^{ S_{n,1}/T}+e^{S_{n,2}/T}$ is
a normalization factor. This formalism is often called the logistic
selection model or logit rule \cite{Anderson}, and it bears some analogy with
the Glauber rate in physical systems, if we
assume that energy plays the role of negative satisfaction.
Suppose that the buyer $n$ chooses the $i$th seller to purchase a product
of freshness $h_{n,i}$ for price $x_{n,i}$ at time $t$.
This product updates
his or her satisfaction from this seller and affects the buyer's next purchase
at $t' > t$. We assume that the satisfaction at $t'$ can be described as
\begin{equation}
S_{n,i}(t') = \alpha S_{n,i}(t) + (1-\alpha)\left\{g\left(1-x_{n,i}^t \right)
+(1-g) h_{n,i}^t \right\},
\label{eq.sa}
\end{equation}
where $\alpha$ characterizes the buyer's memory and $g$ is a {\em greed
factor} to determine the sensitivity to the price.
The superscript $t$ means that the price and freshness are measured at time
$t$.

\section{Analysis}

\subsection{Time evolution of probabilities}
\label{subevol}
Let us assume that the buyers can be represented by a single average
buyer. In other words, we will neglect fluctuations in the buyers'
satisfaction as well as their behavior to drop the buyer index $n$. Then,
the probability to choose the $i$th seller becomes $p_i = Z^{-1} e^{ S_i/T}$
with $Z \equiv \sum_i e^{ S_i/T}$.
At time $t$,
our average buyer pays $\bar{x}_i^t$ to purchase a product of freshness
$\bar{h}_i^t$ from the $i$th seller.
According to Eq.~(\ref{eq.sa}), the buyer's satisfaction evolves as
\begin{equation}
S_i(t+\tau_0) = \alpha S_i(t) + (1-\alpha)\left\{g\left(1-\bar{x}_i^t\right)
+(1-g) \bar{h}_i^t\right\},
\end{equation}
where $\tau_0$ denotes the average time interval to make a purchase.
If the memory decays very slowly, i.e., $0< 1-\alpha \ll 1$, we
may regard $S_i$ as a smooth function of $t$ to obtain
\begin{equation}
\label{eq.sdot}
\dot{S}_i= \frac{1-\alpha}{\tau_0}\left[g \left(1-\bar{x}_i \right)
+(1-g) \bar{h}_i  - S_i \right],
\end{equation}
where the dot means the time derivative. Note also that we have suppressed the
superscript $t$, because Eq.~(\ref{eq.sdot}) is not a difference equation
between $t$ and $t+\tau_0$, but a differential equation at time $t$.
The time evolution of $p_i$ can also be expressed as
\begin{eqnarray}
\dot{p}_i &= \frac{1}{T} \dot{S}_i \frac{e^{ S_i/T}}{Z}
-\left(\sum_j \frac{1}{T} \dot{S}_j
e^{ S_j/T}\right)\frac{e^{ S_i/T}}{Z^2}\nonumber\\
&=\frac{1}{T} p_i \left( \dot{S}_i - \frac{1}{Z}\sum_j \dot{S}_j
e^{ S_j/T} \right)\nonumber\\
&=\frac{1}{T}p_i \left( \dot{S}_i - \langle \dot{S} \rangle \right),
\label{eq.ss}
\end{eqnarray}
where $\langle X \rangle = \sum_j p_j X_j$ means the average over the sellers.
Equation~(\ref{eq.ss}) can be understood as the replicator
equation~\cite{Smith1974,Taylor1978,Hofbauer1979}, where Eq.~(\ref{eq.sdot})
plays the role of fitness.
By substituting Eq.~(\ref{eq.sdot}) into Eq.~(\ref{eq.ss}), we find
\begin{eqnarray}
\dot{p}_i &=\frac{1-\alpha}{\tau_0 T}  p_i
\left\{\left[g(1-\bar{x}_i)+(1-g) \bar{h}_i \right. \right.- \left. \left.S_i  \right]
-\left\langle
g(1-\bar{x}) +(1-g) \bar{h} - S
\right\rangle
\right\}\nonumber\\
&=\frac{1-\alpha}{\tau_0 T}  p_i
\left\{\left[g(1-\bar{x}_i)+(1-g) \bar{h}_i\right.\right.  - \left.\left.T\ln p_i \right]\right. \nonumber\\
&\qquad\qquad\qquad\qquad\qquad\qquad\quad  \left.-\left\langle
g(1-\bar{x}) +(1-g) \bar{h} -  T\ln p
\right\rangle
\right\},
\label{eq.pdot}
\end{eqnarray}
where we have used $S_i = T\left( \ln p_i + \ln Z\right)$.

\subsection{Stationary state}
\label{substat}
We first consider a stationary state in which all $p_i$'s are constant.
The average fraction of products purchased from the $i$th seller during a time
interval $(t,t+dt)$ is expressed as
\begin{equation}
\lambda_i(t) dt = \frac{N_a p_i }{N_p \tau_0}dt,
\end{equation}
because the average number of buyers visiting $i$ per unit time is
$N_a p_i/\tau_0$.
The distribution of product ages $\tau$ at the $i$th seller is denoted as
$\phi_i(\tau,t)$ with a normalization condition $\int_0^\infty d\tau
\phi_i(\tau,t)=1$. The buyer randomly chooses a product, regardless
of its age. The time evolution of $\phi_i (\tau,t)$ is thus written as
\begin{equation}
\phi_i (\tau+dt, t+dt) = \phi_i(\tau, t) -\left[ \lambda_i(t)dt\right]
\phi_i(\tau,t).
\label{eq.phii}
\end{equation}
By expanding Eq.~(\ref{eq.phii}) to the linear order of $dt$, we find that
\begin{equation}
\frac{\partial \phi_i}{\partial \tau}(\tau,t) +
\frac{\partial \phi_i}{\partial t}(\tau,t)
\approx -\lambda_i(t)\phi_i(\tau,t).
\end{equation}
This equation has a stationary solution
$\phi_i^{\text{st}} (\tau) = \lambda_i e^{-\lambda_i \tau}$,
from which $\bar{h}_i$ for the stationary state is obtained as
$\bar{h}_i = \int_0^\infty d\tau \phi_i^{\text{st}} (\tau) h(\tau)
= 1 / \left[1+ (\lambda_i \tau_1)^{-1} \right]$.
If we introduce $R \equiv \frac{N_a\tau_1}{N_p \tau_0}
= \lambda_i \tau_1 / p_i$, this result can also be written as
\begin{equation}
\label{eq.avh}
\bar{h}_i= \left(1+\frac{1}{R p_i}\right)^{-1}.
\end{equation}
In addition,
the value of $\bar{x}_i \equiv \int_0^\infty d\tau \phi_i^{\text{st}} (\tau)
x(\tau)$ can be computed from the stationary distribution as follows:
\begin{equation}
\bar{x}_i = \int_0^\infty d\tau \lambda_i e^{-\lambda_i \tau}
\left[1-\exp\left(-\frac{e^{-\tau/\tau_1}}{h_c}\right)\right].
\end{equation}
Let us introduce another variable $y \equiv e^{-\tau/\tau_1}h_c^{-1}$.
The change of variables then leads to
\begin{equation}
\bar{x}_i =1 -  p_i R h_c^{p_i R} \gamma\left(p_i R, h_c^{-1} \right),
\label{eq.avx}
\end{equation}
where $\gamma(s,x)$ is the lower incomplete gamma function
$\gamma(s,x) \equiv \int_0^x dy  e^{-y}y^{s-1}$.

A possible stationary state is such that each seller is chosen with equal
probability, i.e., $p_i = p_0 \equiv 1/2$. We will check the stability of such
a state under small perturbation in the following way: First, let us rearrange
Eq.~(\ref{eq.pdot}) as
\begin{eqnarray}
\dot{p}_i =\frac{1-\alpha}{\tau_0 T}
\left\{g \left[p_i(1-\bar{x}_i)- p_i\left\langle
1-\bar{x}\right\rangle\right]
+  (1-g) \left[p_i\bar{h}_i - p_i\left\langle\bar{h}\right\rangle\right] \right.  \nonumber\\
\qquad\qquad\qquad\qquad\qquad\qquad\qquad\qquad\quad
 - \left. T\left[p_i\ln p_i-p_i\left\langle\ln p
\right\rangle\right] \right\}.
\label{eq.newpdot}
\end{eqnarray}
By adding small perturbation, we change the probabilities to
$p_i=p_0+\epsilon_i$ with $\sum_j \epsilon_j=0$.
The expressions inside the square brackets on the right-hand side (RHS) of
Eq.~(\ref{eq.newpdot}) can then be expanded as
\begin{eqnarray}\label{eq.term}
p_i(1-\bar{x}_i) -p_i\langle1-\bar{x}\rangle&=
\mathcal{F}(R,p_0, h_c)\epsilon_i+ \mathcal{O}(\epsilon_i^2),\\
p_i \bar{h}_i - p_i\left\langle\bar{h}\right\rangle  &=
\frac{R p_0}{( 1+R p_0)^2}\epsilon_i+ \mathcal{O}(\epsilon_i^2),\\
p_i \ln p_i -p_i\left\langle\ln p
\right\rangle    &=\epsilon_i + \mathcal{O}(\epsilon_i^2),
\end{eqnarray}
with
\begin{equation}
\mathcal{F}(R,p_0, h_c) = \left.   \frac{\partial}{\partial \epsilon}\left\{ p_0(p_0+\epsilon) R h_c^{(p_0 +\epsilon) R} \gamma\left[(p_0+\epsilon)R, h_c^{-1}\right]\right\} \right|_{\epsilon=0}.
\end{equation}
We note that
the zeroth order of $\epsilon_i$ does not exist, so that
Eq.~(\ref{eq.newpdot}) can be rewritten to the linear order of $\epsilon_i$ as
\begin{equation}
\dot{\epsilon}_i \approx \frac{1-\alpha}{\tau_0 T}
\left\{g \mathcal{F}(R,p_0, h_c)
+(1-g)\frac{R p_0}{( 1+R p_0)^2}
\\ - T
\right\}\epsilon_i.
\label{eq.newpdot2}
\end{equation}
The unperturbed stationary state is stable when the prefactor of
$\epsilon_i$ on the RHS is negative. In other words,
we can observe a symmetric phase, in which $p_i=1/2$ for every $i$,
provided that
\begin{equation}
T > T_c(g) \equiv g\mathcal{F}(R,p_0, h_c)+(1-g)\frac{R p_0}{( 1+R p_0)^2},
\label{eq.tc}
\end{equation}
for given $R$ and $q$.
One should note that multiple stable states can coexist in some range
of $R$~\cite{Lambert2011}. If $R$ is small enough, however, we may say that
the symmetric state is the only possibility when $T > T_c$.
The question is what happens when $T$ becomes lower than $T_c$.
When the greed factor $g$ approaches zero, the above analysis reduces
to the result of Ref.~\cite{Lambert2011}, in which an asymmetric phase
emerges below $T_c(g=0)$. The origin of the asymmetric phase can be explained
as follows:
If the buyer's choice gets slightly biased against a certain seller
by chance, the products of this seller get older, which in turn
lowers the probability to choose this seller, forming a positive feedback loop.
In the end, only a single seller occupies the whole market, which is a stable
stationary state at low $T$.
Even if $g$ has a small finite value, it is reasonable
to expect the same spontaneous symmetry-breaking phenomenon.
When $g$ gets high enough, however, the price system comes into play in a
nontrivial way as will be explored below.

\subsection{Oscillatory phase}\label{subosc}

\begin{figure}
\center\includegraphics[width=0.60\textwidth]{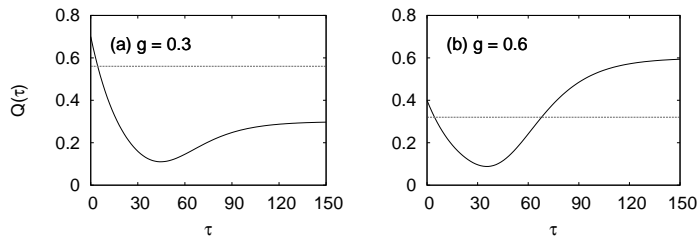}
\caption{$Q(\tau)$ for (a) $g=0.3$ and (b) $g=0.6$. The other parameter
values are as given in Sec.~\ref{numeric}.
The dotted horizontal lines represent $Q_0$,
and the existence of the second crossing point between $Q(\tau)$ and $Q_0$
distinguishes the oscillatory phase from the asymmetric phase (see text).}
\label{fig:q}
\end{figure}

Let us rewrite Eq.~(\ref{eq.sdot}) as
\begin{equation}
\dot{S}_i = \frac{1-\alpha}{\tau_0}\left[-S_i + Q_i(\tau) \right],
\label{eq.sit}
\end{equation}
by defining $Q_i(\tau) \equiv g [1-\bar{x}_i(\tau)]+(1-g)\bar{h}_i(\tau)$.
It implies that we may expect $S_i \approx Q_i (\tau)$ when the system is
evolving slowly. Suppose that only one seller, say, $i=1$, is selected due to
the symmetry-breaking mechanism explained above. This seller is always equipped
with new products, so the satisfaction from this seller will be $S_1 \approx Q_1
(0)$. On the other hand, $\tau$ increases for the other seller as time $t$
goes by, so $S_2(t)$ will follow the trajectory of $Q_2(\tau)$.
Figure~\ref{fig:q} shows  two possible cases of $Q_i (\tau)$, i.e., with
small and large greed factors, respectively. When $g$ is small,
$Q_2 (\tau > 0)$ stays below $Q_1(0)$, so the asymmetric phase remains
stable [Fig.~\ref{fig:q}(a)]. On the other hand, if $g$ is large enough,
the low price compensates for the freshness reduction, and $Q_i(\tau)$ can
actually be greater than $Q_i(0)$ when $\tau$ exceeds a certain threshold
[Fig.~\ref{fig:q}(b)]. Now, the second seller begins to be selected, and the
situation is less advantageous to the first seller: The value of $Q_1$
decreases,
because some products are not sold, which lowers the freshness further.
On the other hand, $Q_2$ does not decrease, because old products make positive
contributions to $Q_1$ due to their low prices while refreshed products
also attract buyers. The result is that the second seller is preferred
until the prices of the other seller become low enough. To sum up,
they take turns in playing the role of the champion and the challenger,
which is the origin of the oscillating behavior.

The above argument leads to some quantitative predictions:
Assume that the behavior can be
approximated as that of the asymmetric phase on a short time scale.
We therefore suppose that $p_i=1$ so that only seller $i$ is being selected.
In this case, $\bar{h}$ and $\bar{x}$ are derived in Eqs.~(\ref{eq.avh})
and (\ref{eq.avx}), respectively, whereby we obtain an explicit expression for
$Q_i$ at $p_i = 1$. This is denoted as $Q_0$ and represented as dotted
horizontal lines in Fig.~\ref{fig:q}. Note that $Q_0 \approx Q_i(\tau=0)$,
because, by assumption, this sellers' products are being replaced with fresh
ones. However, it takes some finite time to sell the products, so $Q_0$ is
slightly lower than $Q_i(\tau=0)$. As a result, $Q_i(\tau)$ has a trivial
crossing at $\tau \sim O(N_p \tau_0 / N_a)$ (Fig.~\ref{fig:q}).
For old products to be as competitive as fresh ones, $Q_i(\tau)$ should
have another crossing with $Q_0$ at some $\tau = \tau^\ast \gg O(N_p \tau_0
/ N_a)$.
The shape of the curves in Fig.~\ref{fig:q} suggests that this condition
can be rephrased as $Q_i (\tau \rightarrow \infty) = g > Q_0$.
It is clear that $Q_i(\tau)$ approaches $g$ as $\tau$ grows,
so that the critical value $g_c$ is a solution of the following equation:
\begin{equation}\label{eq.gc}
g\left[1 -  R h_c^R \gamma\left(R, 1/h_c\right)\right] +
(1-g)\left(1+\frac{1}{R}\right)^{-1} = g,
\end{equation}
where the left-hand side expresses $Q_0$.
We may expect oscillatory behavior when $g$ exceeds $g_c$. At the same time, we
stress that this calculation assumes deterministic behavior when it comes to
buyers, so that the prediction will be the most accurate at $T=0$.
Furthermore, the crossing point $\tau^\ast$ between $Q_i(\tau)$ and
$Q_0$ provides an estimate of the half-period of oscillation. It
is obtained by numerically solving the following equation:
\begin{equation}
g\exp\left(-h_c^{-1} e^{-\tau^*/\tau_1}\right)+(1-g)e^{-\tau^*/\tau_1} = Q_0.
\label{eq.tauprime}
\end{equation}

\section{Numerical results}
\label{numeric}

\begin{figure}
\center\includegraphics[width=0.75\textwidth]{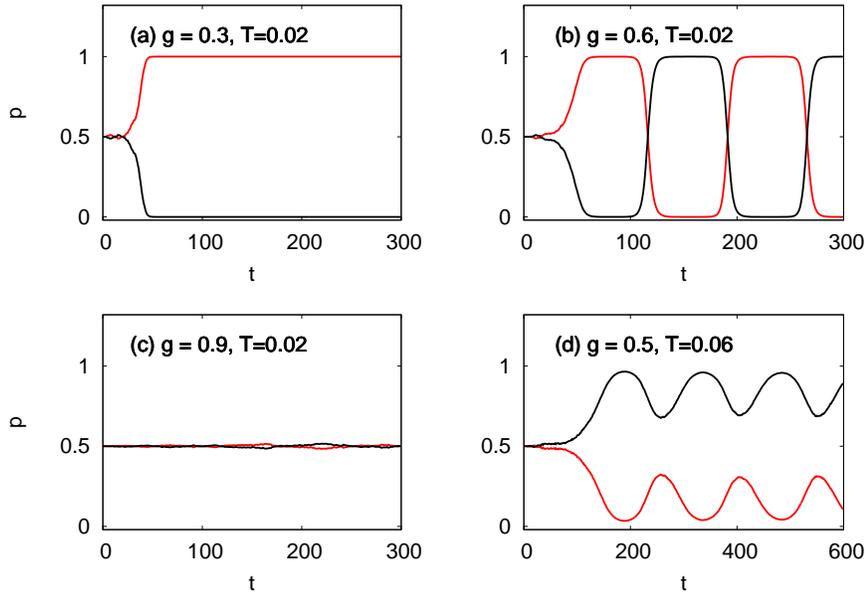}
\caption{Time evolution of the probability to choose each seller (red or black line),
started from an initial condition $p_i = 1/2$. As the greed factor $g$ varies at
a sufficiently low temperature $T=2 \times 10^{-2}$, we observe (a) asymmetric,
(b) oscillatory, and (c) symmetric phases, respectively. (d) With a relatively
high temperature ($T=6 \times 10^{-2}$), we can also find asymmetric
oscillation.}
\label{fig:phases}
\end{figure}

Our numerical simulation shows the existence of the symmetric,
oscillatory, and asymmetric phases as predicted above (Fig.~\ref{fig:phases}).
In this simulation, we set the number of products at each seller as
$N_p = 5 \times 10^3$ and the number of buyers as $N_a = 10^2$. The typical
time interval of a purchase is given as $\tau_0 = 10^{-1}$, and freshness is
assumed to decay with a characteristic time scale $\tau_1 = 20$. The price
policy is parametrized by a freshness threshold $h_c = 5 \times 10^{-2}$,
and the memory factor is set as $\alpha = 0.99$. With all these parameters
fixed, we change $g$ and $T$ to locate the phase boundaries.
Recall that the boundary of the symmetric phase is predicted
by Eq.~(\ref{eq.tc}), whereas the boundary between the asymmetric and
oscillatory phases is predicted to be $g = g_c$, obtained by solving
Eq.~(\ref{eq.gc}).

\begin{figure}
\center\includegraphics[width=0.5\textwidth]{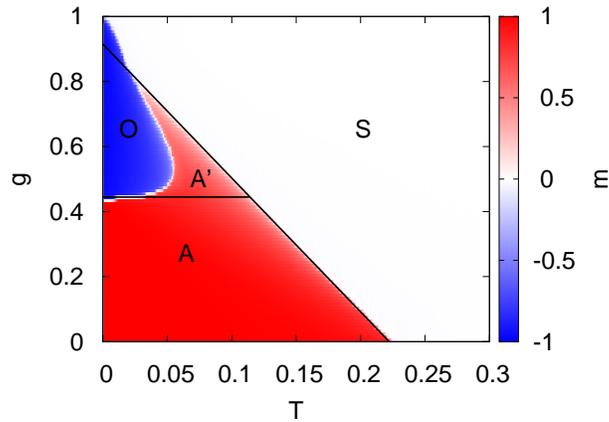}
\caption{
Phase diagram from numerical simulation (see text).
For comparison, the phase boundaries
predicted in Eqs.~(\ref{eq.tc}) and (\ref{eq.gc}) are represented by the solid
lines.  The labels S, A, and O mean the symmetric ($m \approx 0$), asymmetric ($m>0$), and oscillatory ($m<0$)
phases, respectively.
In the region labeled as A', we observe asymmetric oscillation
[Fig.~\ref{fig:phases}(d)].
}
\label{fig:diag}
\end{figure}

Let us define some numerical quantities to distinguish the three phases.
First, the symmetry breaking can be detected by the following:
\begin{equation}
m_A = |\langle p_1(t)-p_2(t) \rangle_t|,
\end{equation}
where $p_i(t) = N_a^{-1} \sum_n p_{n,i}$ is the probability to choose
seller $i$ at time $t$ and $\langle ... \rangle_t$ means the average over time.
Clearly, this quantity will be nonzero only in the asymmetric phase.
To make a distinction between the symmetric and oscillatory phases, we also
measure the following:
\begin{equation}
m_O = \langle|p_1(t)-p_2(t) |\rangle_t-|\langle p_1(t)-p_2(t) \rangle_t|,
\end{equation}
which will be nonzero only in the oscillatory phase. If we combine these two
into a single parameter $m = m_A- m_O$, its value will be positive (negative)
in the asymmetric (oscillatory) phase, and close to zero in the symmetric phase.
Figure~\ref{fig:diag} shows the results: The solid lines are the phase
boundaries estimated by Eqs.~(\ref{eq.tc}) and (\ref{eq.gc}).
The boundary between the symmetric and asymmetric phases
shows an excellent agreement, especially for $g \lesssim 0.8$.
The division between the oscillatory and asymmetric phases is also
consistent with the estimation of $g_c$ when $T$ is low. The region A'
is characterized by asymmetric oscillation, which has not been captured
by our analysis. Such a phase can exist at sufficiently high $T$, because
the unfavored seller sells its old products too early due to buyers'
stochastic choices.

\begin{figure}
\center\includegraphics[width=0.5\textwidth]{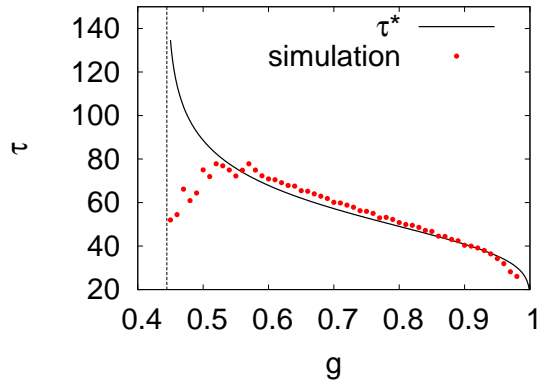}
\caption{
$\tau^\ast$ in Eq.~(\ref{eq.tauprime}) and numerically estimated half-periods,
for different values of $g$ at $T=10^{-3}$. The other parameters are the
same as used in Sec.~\ref{numeric}. Equation~(\ref{eq.tauprime}) gives a good
prediction for the half-periods when $g$ is large enough, but the estimate
becomes inaccurate as $g$ approaches $g_c\approx 0.44$, which is represented by
the dotted vertical line.
}
\label{fig:tau}
\end{figure}

Finally, we compare $\tau^\ast$ with the half-period in the oscillatory phase
in Fig.~\ref{fig:tau}. We choose very low temperature, $T=10^{-3}$, because
the estimation of $\tau^\ast$ can easily be disturbed by noise. As shown in
Fig.~\ref{fig:tau}, the agreement is striking for large $g$. The failure
at small $g$ is related to one of our fundamental assumptions: Recall
that we have assumed that all the buyers are homogeneous by using a
single average buyer in Sec.~\ref{subevol}. When $g$ is close to
$g_c\approx0.44$, this assumption breaks down and the buyers have
different probabilities to choose the sellers. This inhomogeneity dramatically
reduces the period of oscillation.

\section{Discussion}
\label{discuss}

The phase diagram in Fig.~\ref{fig:diag} shows that the price system indeed
restores the symmetry in a broader range of $T$ as compared to the case
without price formally obtained as $g=0$. However, the asymmetric
phase can exhibit oscillations when buyers are sensitive to the price.
One might point out that
our model bears similarity to the Bertrand duopoly, in which two sellers sell
homogeneous products without any possibility of collusion~\cite{maskin}.
The sellers in the Bertrand duopoly compete by setting prices simultaneously,
and the one with a lower price occupies the whole market. If the two sellers
charge the same price, they will evenly share the market. The pure-strategy Nash
equilibrium turns out to be such that both of the sellers set the price equal to
marginal cost of production, which means that they earn nothing from their
business~\cite{tirole}.
In the Bertrand duopoly, therefore, the existence of two sellers is
already enough to actualize the lowest possible price. However, one should note
its underlying assumption that a single seller can cover the whole market.
If this is not the case, e.g., due to limited capacity of production, we
encounter the Edgeworth paradox, which means that this game has no
pure-strategy Nash equilibrium, hence no stationary price. For example, one of
the two firms can make profits by deviating from the marginal cost, because the
other firm alone cannot meet the demand. Although a mixed strategy can
constitute a Nash equilibrium~\cite{dasgupta,vives}, it is hardly feasible in
practice and one would instead observe the Edgeworth price cycle
(Reference~\cite{maskin} also points out that the capacity constraint may
not be a necessary condition for the cycle). The Edgeworth paradox suggests
that the price mechanism can introduce instabilities. However, the cycle is
actually a pseudo-dynamic process in the sense that it originates from an
equilibrium concept.
In contrast, our oscillatory phase is born out of dynamical rules without
any consideration of strategic equilibrium.

To consider some strategic aspects in our model,
let us suppose that the variables $T$ and $g$ characterizing buyers are
just given parameters when viewed from the sellers. The sellers can instead
decide $h_c$ to control the price policy, and the symmetry between the sellers
suggests that they will end up with the same $h_c$.
The critical temperature $T_c$ in Eq.~(\ref{eq.tc}) depends on $h_c$ unless
$g=0$. For example, if $h_c \rightarrow 0$, the price will become insensitive to
freshness. It implies that $\mathcal{F}(R, p_0, h_c)$ vanishes in
Eq.~(\ref{eq.tc}) so that the phase boundary of the symmetric phase in
Fig.~\ref{fig:diag} will converge to a line connecting $(T_c(0),0)$ and $(0,1)$
on the $(T, g)$ plane. In a price war, the sellers will
increase $h_c$ to lower the average price, but the possible range of $h_c$ must
be bounded due to the cost of production. We may regard our $h_c$ in the
previous sections as the highest possible one determined by such a competitive
process.  Differently from the Bertrand duopoly, however, the sellers do not
always divide the market half and half, when the mechanism of Lambert et al. is
at work~\cite{Lambert2011}: The market of perishable goods tends to become
monopolistic, especially when the purchasing behavior is deterministic with low
$T$. Our point is that this tendency is nevertheless weakened by the price
sensitivity, either by extending the symmetric phase, or by restoring the
symmetry over a long period in the oscillatory phase.

\section{Summary}
To summarize, we have incorporated a simple price system coupled with
freshness into the duopoly model suggested by Lambert et al. In addition to the
effective temperature $T$ to control the randomness in purchasing behavior, we
have introduced the greed parameter $g$ to determine the sensitivity of
satisfaction to the price.
We have identified the symmetric, asymmetric, and oscillatory phases and
estimated their boundaries in the $(T,g)$ plane. Our numerical simulations
show nice agreements with our analytic results. Based on our analysis, we
conclude that the price system resists the tendency to a monopoly: On one hand,
it lowers the critical temperature $T_c$ below which the symmetric phase becomes
unstable. On the other hand, the market with high $g$ can keep oscillating
without settling on a single seller, preserving the symmetry in a time-averaged
sense.

\ack
S.K.B. gratefully acknowledges discussions with Kyung Pill Kim.
S.D.Y. was supported by Basic Science Research
Program through the National Research Foundation of
Korea funded by the Ministry of Education (Grant No. NRF-2014R1A6A3A01059435).
S.K.B. was supported by Basic Science Research Program through the
National Research Foundation of Korea funded by the Ministry of Science,
ICT and Future Planning (Grant No. NRF-2014R1A1A1003304).
S.K.B. and S.D.Y. thank the LIPHY laboratory (CNRS and Univ.~Joseph Fourier) in
Grenoble for its hospitality during the initial stages of this work.

\section*{References}
\bibliography{shop}

\end{document}